\documentclass[aps,prd,twocolumn]{revtex4}[12pt]
\usepackage[latin1]{inputenc}
\usepackage{amsmath}
\usepackage{amssymb}
\usepackage[dvips]{graphicx,epsfig}

\input{epsf}

\def \beq  {\begin{equation}}
\def \eeq  {\end{equation}}
\def \ber  {\begin{eqnarray}}
\def \eer  {\end{eqnarray}}

\begin{document}
\title{Solar System Constraints on Scalar Tensor Theories with Non-Standard Action}
\author{N. Chandrachani Devi$^1$, Sudhakar Panda$^{2,3}$,  Anjan A Sen$^1$}
\affiliation{$^1$ Centre of Theoretical Physics, Jamia Millia Islamia, New Delhi-110025, India}
\affiliation{$^2$ Harish-Chandra Research Institute, Chhatnag Road,
Jhusi, Allahabad-211019, India}
\affiliation{$^3$ Erskine Fellow, Department of Physics and Astronomy, University of Canterbury, 
Christchurch, New Zealand}
\date{\today}

\begin{abstract}  
 We compute the Post-Newtonian parameter (PPN),$\gamma$, for scalar-tensor gravity theory when the action functional for the scalar field is a non-standard one, namely the Dirac-Born-Infeld (DBI) type action, used in the literature for a tachyon field. We investigate two different cases (Linear and conformal couplings) when the scalar field is non-minimally coupled to gravity via the scalar curvature. We find that the PPN parameter $\gamma$, which measures the amount of space curvature per unit rest mass, becomes a function of the effective mass of the scalar field. Using this PPN parameter, we calculate the time delay $\Delta\tau$ for the signal to travel the round trip distance bertween  a ground based antenna and a reflector placed in a spacecraft which is produced due to the grvitational field of Sun. We use this $\Delta\tau$ to compare our result with that obtained by the Cassini mission and derive the constraints on the model parameters.
\end{abstract}

%
\maketitle

Some important Solar System tests for theories of relativistic gravity include the gravitational redshift, the deflection of light by the Sun, the precession of the perihelion of a planetary orbit, etc. Einstein's general relativity  is consistent with these experimental tests. While considering these tests, it is useful to have a framework in which the predictions of different theories are parametrized in a systematic way. The parametrized-Post-Newtonian (PPN) framework \cite{will} has become a basic tool to connect gravitational theories with these experiments.
In PPN formalism, one takes the slow-motion, weak-field limit (the post Newtonian limit) and expands the space time metric $g_{\mu\nu}$ of any gravitational theory about the Minskowski metric $\eta_{\mu\nu}=dia(-1,1,1,1)$  in terms of  Newtonian potentials and thereby obtains the post Newtonian corrections by comparing with the standard expansion of the metric in terms of the PPN parameters (the coefficients of the Newtonian potential). These parameters may be different in different theories of gravity. For general relativity the values are, $\gamma=1$ and $\beta=1$ (where $\gamma$ measures the amount of space time curvature per unit mass and $\beta$ represent the amount of non-linearity in the superposition law of gravity). Restricting to the case of point sources, these formalism is often called Eddington-Robertson-Schiff formalism and parameters are termed as Eddington parameters ( In the rest of the paper, we call them in general the PPN parameters).   

The recent renewed interests in scalar fields, minimally coupled to gravity, originates from their role in describing two accelerating phases in the history of the expanding universe: one during very early time associated with very high energy scales\cite{guth} and another during much later period (more precisely at present epoch) with much lower energy scale \cite{de}. During both of these epochs, the slowly varying scalar field can mimic an effective cosmological constant which in turn can violate the strong energy condition resulting in the accelerated expanding phase.

Scalar tensor theories are generalization of these minimally coupled field theories in a sense that the scalar fields are now non minimally coupled with the gravity sector. Although a long range scalar field, as a gravitational field, was first introduced by Jordan\cite{jordan}, the standard example most studied in the literature is the Brans-Dicke (BD) theory \cite{Brans(1961)} and the analysis of the PPN parameter, in this context is carried out in \cite{ppn}. Another class of scalar tensor theories  naturally arises in  Superstring theory where one encounters the dilaton field non minimally coupled with the gravity sector. The analysis for the PPN parameter for such gravity-dilaton system has been performed in \cite{Kalyana Rama(1994)}. The reason that these scalar tensor theories are one of the most natural alternatives to the general relativity (GR) is due to the fact that they respect most of the symmetries in GR like local Lorentz invariance, energy momentum conservation etc. The recent relevance of the scalar tensor theories is also due to the fact that the modified gravity theories like $f(R)$ \cite{fr}, where one adds nonlinear contribution of the curvature scalar $R$ in the gravity action to explain the late time acceleration of the universe, can be described as an effective scalar tensor theory. PPN parameters for such $f(R)$ gravity theories have also been studied by using the dynamical equivalence between $f (R)$ and scalar-tensor theory gravity 
\cite{Olmo:2005jd},\cite{Capone:2009xk}. 

In recent years, an alternate possibility of having scalar fields governed by a non-standard action with the Lagrangian density of the form: ${\cal L} = - V(\Phi)\sqrt{1 + X}$, where $V$ is the potential function of the field and $X= \frac{1}{2}\partial_{\mu}\Phi\partial^{\mu}\Phi$, has been proposed \cite{Sen(2002)Bergshoeff et al.(2000)Garousi(2000)Kluso(2000)}. This is a generalization of the Dirac-Born-Infeld (DBI) action and captures the dynamics of the scalar field (called tchyon) living on the world volume of a non-BPS brane in Type II string theory \cite{Sen(2005)}. This non-standard form of the action, as argued in \cite{Padmanabhan(2002)}, is another form for the action functional of a relativistic scalar field. Such an action has attracted significant attention in the context of both inflation\cite{dbi}  as well as late time acceleration of the universe \cite{dbilt}. In fact such DBI action with the scalar field being non-minimally coupled to gravity has been studied in the context of inflation \cite{CDPO}.

In spite of the fact that such a new class of scalar tensor theories explain some interesting cosmological observations, they are expected to be severely constrained by the local gravity tests, e.g, Solar System experiments which give very stringent bound on any deviations from standard GR.  In this note, we carry out our analysis of such an attempt.  


We start with the non-standard DBI  form of the action  for the scalar field $\Phi$ which is non minimally coupled to gravity in the following form: 
\begin{widetext}
 \beq
S=\frac{1}{16\pi G}\int d^4x \sqrt{-g}
\Bigl(F(\Phi)~R -
\tau_{3}V(\Phi)\sqrt{1+\eta^2~g^{\mu\nu}
\partial_{\mu}\Phi
\partial_{\nu}\Phi}\Bigr)
+ S_m[\psi_m; g_{\mu\nu}]\ 
\label{action}
\eeq
\end{widetext}
where $G$ is the bare gravitational constant, $R$ is the scalar curvature of the metric $g_{\mu\nu}$ and $S_{m}$ is the action for non-relativistic matter field.  This matter field does not evolve but contributes to the dynamical equation for the metric field through the energy momentum tensor. The two parameters $\eta$ and $\tau_3$ are introduced to keep track of the dimensions.  $\eta$ has the dimension of $[length]$ and $\tau_3$ has dimension of $[length]^{-2}$. In the context of string theory they are string length and the tension of a non-BPS D3-brane respectively. The field $\Phi$ in this case is dimensionless. However, as alluded earlier, in this analysis since we will not invoke the dynamics in the context of string theory, rather we simply consider the case of a scalar field with non-standard action, we consider $\eta$ and $\tau_3$ as simply some dimensional constants. Similarly, the potential function, at this stage is a smooth function and otherwise is arbitrary. We will have more comments on this at a later stage of our analysis. Note that the non-minimal coupling of the scalar field with gravity is specified by an arbitrary function $F(\Phi)$.

The dynamical equation for the metric field $g_{\mu\nu}$ which is obtained by variation of the action equation (\ref{action}) with respect to the metric tensor $g_{\mu\nu}$ is of the form:
\begin{widetext}
 \beq
F(\Phi) \left(R_{\mu\nu}-\frac{1}{2}g_{\mu\nu}R\right)
= 8\pi G T_{\mu\nu}
- \tau_3\frac{V(\Phi)}{2} \left(g_{\mu\nu}\sqrt{1+\eta^2~g^{\rho\sigma}
\partial_{\rho}\Phi
\partial_{\sigma}\Phi}-\frac{\eta^2\partial_{\mu}\Phi\partial_{\nu}\Phi}{\sqrt{1+\eta^2~g^{\rho\sigma}
\partial_{\rho}\Phi
\partial_{\sigma}\Phi}}
\right)+\nabla_\mu\partial_\nu F(\Phi) - g_{\mu\nu} \Box F(\Phi)
\label{Gmu}
\eeq
\end{widetext}
Similarly, the dynamical equation for the  scalar field can be obtained by varying the action with respect to this field. However, due to the nature of non-minimal coupling, this variation leads to the presence of the scalar curvature in the dynamical equation. We eliminate this curvature scalar using the above equation and present the resulting equation below:

\begin{widetext}
\begin{align}
& -\frac{\tau_3 V(\Phi)\eta^2 \Box \Phi}{\sqrt{1+\eta^2~g^{\mu\nu}
\partial_{\mu}\Phi \partial_{\nu}\Phi}}
-\frac{\tau_3 V'(\Phi)\eta^2\partial_{\mu}\Phi\partial^{\mu}\Phi}
{\sqrt{1+\eta^2~g^{\rho\sigma} \partial_{\rho}\Phi \partial_{\sigma}\Phi}}+\frac{\tau_3 \eta^2 V(\Phi) g^{\mu
\alpha}(\partial_{\alpha}\Phi)(\nabla_{\mu}
\partial_{\beta}\Phi)(\partial^{\beta}\Phi)}{(1+\eta^2~g^{\rho\nu}
\partial_{\rho}\Phi \partial_{\nu}\Phi)^{3/2}} 
 +8\pi GT\frac{F'(\Phi)}{F(\Phi)}\nonumber\\
 & -2\tau_3 V(\Phi)\frac{F'(\Phi){\sqrt{1+\eta^2~g^{\mu\nu} \partial_{\mu}\Phi
\partial_{\nu}\Phi}}}{F(\Phi)}
+\frac{\tau_3 \eta^2 V(\Phi)\partial_{\rho}\Phi \partial^{\rho}\Phi
 F'(\Phi)}{2 F(\Phi)\sqrt{1+ \eta^2~g^{\mu\nu}\partial_{\mu}\Phi\partial_{\nu}\Phi}} 
-\frac{3F'(\Phi)\Box F(\Phi)}{F(\Phi)} 
 +\tau_3 V'(\Phi){\sqrt{1+\eta^2~g^ {\mu\nu} \partial_{\mu}\Phi \partial_{\nu}\Phi}} = 0
 \label{field}
\end{align}
\end{widetext}

In the above equations $T_{\mu\nu}$ is the energy momentum tensor for the matter field contributed from the matter action $S_m$ which is taken in the form $ T_{\mu\nu}= diag(\rho,p,p,p)$ . $T$ denotes the trace of it. Prime here denotes the differentiation with respect to the field $\Phi$. In what follows, we restrict to  two different forms for the coupling function $F(\Phi)$ for  subsequent analysis.

\section{Case(i): $\!\!  F(\Phi)=\Phi$}

We specify  the non-minimal coupling to be linear in the field and expand the field and the metric  around a constant uniform background field $\Phi_0$ and a Minkowski metric $\eta_{\mu\nu}= diag(-1,1,1,1)$:
\beq
\Phi  =  \Phi_0 + \psi
\eeq
\beq
g_{\mu\nu} = \eta_{\mu\nu} +h_{\mu\nu},
\eeq

\noindent
where $h_{\mu\nu} << 1$ and $\psi << \Phi_{0}$.  Here $\psi$ represents the local deviation from $\Phi_0$.
Similarly, we also expand $V(\Phi)$ as $V(\Phi)=V(\Phi_0)+\psi V'(\Phi_0)+\frac{\psi^2}{2}V''(\Phi_0)$ as well as $V'(\Phi)$ as $V'(\Phi)=V'(\Phi_0)+\psi V''(\Phi_0)$.  We also assume that $\Phi_{0}$ is the location of extrema of the potential and hence $V'(\Phi_{0}) = 0$ for our subsequent calculations. We assume that at this extrema, $V(\Phi)$ has a very small  but nonzero value, i.e $V(\Phi_{0}) = \epsilon, \hspace{1mm} \epsilon << 1$, thus assuming the existence of a small but non zero constant scalar energy density.  

The dynamical equations (\ref{Gmu}) and (\ref{field}) thus reduce to equations in terms of $\psi$ and $h_{\mu\nu}$ where we can neglect all the higher order terms involving $\psi$ and $h_{\mu\nu}$ and $\epsilon$. For further simplification, we choose the following gauge:

\beq
h^\mu_\nu,_\mu-\frac{1}{2}h^\mu_\mu,_\nu=\frac{1}{\Phi_0}\psi,_\nu.
\eeq

\noindent
We also define a new variable $\chi_{\mu\nu}=h_{\mu\nu}-\eta_{\mu\nu}\frac{h}{2}-\eta_{\mu\nu} \frac{\psi}{\Phi_0}$, where $h=h_\mu^\mu$.

Working in the above gauge and in terms of the new variable $\chi_{\mu\nu}$, equations (2) and (3) now become:

\ber
-\frac{\Phi_0}{2}\Box \chi_{\mu\nu} = 8 \pi G T_{\mu\nu}
-\frac{ \tau_3 V(\Phi_0)\eta_{\mu\nu}}{2}\nonumber\\
-\frac{ \tau_3 V(\Phi_0)h_{\mu\nu}}{2}
-\frac{ \tau_3 V'(\Phi_0)\eta_{\mu\nu}\psi}{2}
\label{chi},
\eer

\beq
\Box \psi-\frac{ \tau_3 V''(\Phi_0)\Phi_0}{\left(3+\tau_3 \eta^2 \epsilon \Phi_0\right)}\psi = \frac{8\pi GT}{\left(3+\tau_3 \eta^2 \epsilon \Phi_0\right)}-\frac{2\tau_3 \epsilon}{\left(3+\tau_3 \eta^2 \epsilon \Phi_0\right)}
\label{psi}.
\eeq

\noindent
We are interested in gravitational field around astrophysical objects like sun or our earth. In such situations, the gravitational field is approximately static and hence we ignore time derivative in the equations of motion and we also set the pressure, $p\simeq 0$ as the systems being non relativistic. The equations (\ref{chi}), (\ref{psi}) then reduce to
(we assume $V(\Phi_0)=0$ and ignore all 2nd and higher order terms involving $\epsilon, h_{\mu\nu}$ and $\psi$)
\beq
\nabla^2 \psi-\frac{ \tau_3 V''(\Phi_0)\Phi_0}{\left(3+\tau_3 \eta^2 \epsilon \Phi_0\right)}\psi = -\frac{8\pi G\rho}{\left(3+\tau_3 \eta^2 \epsilon \Phi_0\right)}-\frac{2\tau_3 \epsilon}{\left(3+\tau_3 \eta^2 \epsilon \Phi_0\right)}
\eeq
\beq
\nabla^2\chi_{00} = -\frac{16\pi G\rho}{\Phi_0}-\frac{\tau_3 \epsilon}{\Phi_0}
\eeq
\beq
\nabla^2\chi_{ij} = \frac{\tau_3 \epsilon \delta_{ij}}{\Phi_0}
\eeq

\noindent
We set the energy density, $\rho=M_s \delta(r)$ where we assume the presence of a source at the origin $r=0$. Since in the solar system the Sun represents the main contribution to the matter energy density, $ M_s$ is the Newtonian mass of the Sun. With this, we obtain the solutions as,
\begin{align}
\psi(r) = & \frac{2G M_s}{r\left(3+\tau_3 \eta^2 \epsilon \Phi_0\right)} \exp\left[-\sqrt{\frac{V''(\Phi_0) \tau_3 \Phi_0}{\left(3+\tau_3 \eta^2 \epsilon \Phi_0\right)}}r\right] +\nonumber\\
& \frac{2\epsilon}{V''(\Phi_0) \Phi_0}
\end{align}
\beq
\chi_{00}= \frac{4GM_s}{\Phi_0 r} - \frac{\tau_3 \epsilon}{6 \Phi_0}r^2
\eeq
\beq
\chi_{ij}= \frac{\tau_3 \epsilon \delta_{ij}}{6 \Phi_0}r^2
\eeq
The solutions for $h_{\mu\nu}$ are
\begin{align}
h_{00}= & \frac{2GM_s}{\Phi_0 r} \Bigl(1+\frac{ \exp\left[-\sqrt{\frac{\tau_3 \Phi_0}{\left(3+\tau_3 \eta^2 \epsilon \Phi_0\right)}}m_{eff} r\right]}{\left(3+\tau_3 \eta^2 \epsilon \Phi_0\right)}\Bigr)\nonumber\\
& +\frac{\tau_3 \epsilon}{6 \Phi_0}r^2+\frac{2\epsilon}{m_{eff}^2\Phi_0^2}
\end{align}
\begin{align}
h_{ij}= & \frac{2GM_s\delta_{ij}}{\Phi_0 r} \Bigl(1-\frac{\exp\left[-\sqrt{\frac{\tau_3 \Phi_0}{\left(3+\tau_3 \eta^2 \epsilon \Phi_0\right)}}m_{eff} r\right]}{\left(3+\tau_3 \eta^2 \epsilon \Phi_0\right)}\Bigr)\nonumber\\
& -\frac{\tau_3 \epsilon}{6 \Phi_0}\delta_{ij}r^2 - \frac{2\epsilon}{m_{eff}^2\Phi_0^2}\delta_{ij}.
\end{align}

\noindent
Here, we have defined the ``effective mass" of the scalar field as $m_{eff}^2=V''(\Phi_0)$.  The second term inside the first bracket in the right hand side in both equations above is the correction to the standard term $\frac{2G M_{s}}{r}$. This is similar to the correction obtained by Perivolaropoulos for a massive Brans-Dicke theory  \cite{Perivolaropoulos(2010)} where the term $\tau_{3}\eta^{2}\epsilon\Phi_{0}$ and $\sqrt{\tau_{3}}m_{eff}$ play the similar role as $2\omega$ and $m$ for the massive Brans-Dicke case.  In our model, $\tau_{3}$ and $\eta$ should be nonzero in order to study the effect of the nonstandard action. Hence the $\omega = 0$ in the massive Brans-Dicke case corresponds to $\epsilon = 0$ in our case. In this case, we recover the exactly same solutions for $h_{\mu\nu}$ as one obtains for a massive Brans-Dicke case with $\omega = 0$. Given the fact that $\epsilon$ represents the nonzero value of the potential at its extrema, one observes that for potentials with $V(\Phi_{0}) = 0 $ with $\Phi_{0}$ being the location of the extrema, scalar field with DBI type action produces the same gravitational field under weak field approximation as one gets for massive Brans-Dicke gravity with similar potential but with vanishing BD parameter $\omega$. 

But for nonzero $\epsilon$, one gets two additional terms, one varies as $r^2$ and other one is a constant.  The constant term can be removed with a suitable coordinate transformation $x^{\mu} \rightarrow x^{\mu} + \zeta^{\mu}$ where $\zeta^{\mu} = \frac{\epsilon}{m_{eff}^2 \Phi_{0}^2} x^{\mu}$. With this coordinate transformation, the gauge condition (6) that we have used to solve the system, remains invariant and hence this is an artifact of a gauge mode.


The term involving $r^2$ is similar to what one gets in the presence of a cosmological constant. This term can not be  removed by any coordinate transformation. It is due to the nonzero $\epsilon$ which is the value of the potential at its extremum around which we expand it. In the Einstein's equation (2), this appears as an effective cosmological constant term which is essentially the scalar field constant energy density.  Although in the solutions for $h_{00}$ and $h_{ij}$, this term gives a $r^2$ contribution, typical for a cosmological constant, but the solutions does not reduce to the de-Sitter solution for $M_{s}=0$ or $r\rightarrow \infty$. This is because the Einstein's equation (2) is different from that for a de-Sitter Universe due to the presence of the nontrivial scalar field and its derivatives (for similar solution see \cite{Olmo:2005jd}).


Now, using the standard expansion of the metric in terms of  $\gamma$, the Post-Newtonian parameter 
\beq
 g_{00} =-1+2u 
\eeq
\beq
g_{ij} = (1+2\gamma u)\delta_{ij}
\eeq
 where $u$ is the Newtonian potential, we rewrite expressions of $h_{00}$ and $h_{ij}$, omitting the $r^2$ term, as
\beq
 h_{00}= \frac{2G_{eff} M_s}{r} ,
\eeq 
\beq
 h_{ij}= \delta_{ij}\left(\frac{2G_{eff}\gamma M_s}{r}\right) .
\eeq

\noindent
where  the effective Newton's constant, $G_{eff}$  is defined to be
\beq
 G_{eff}= \frac{G}{\Phi_0}\Bigl(1+\frac{\exp\left[-m_{1} r\right]}{\left(3+2\omega\right)} 
 +\frac{\epsilon\tau_{3} r^3}{12GM_s}\Bigr)
\eeq 

\noindent
where $2\omega =\tau_{3} \eta^2 \epsilon \Phi_{0}$ and  $m_{1} =\sqrt{\frac{\tau_{3}\Phi_{0} m_{eff}^2}{(3+2\omega)}}$ . 

With this, we can write the PPN parameter, $\gamma$ as

\beq
\gamma= 
\frac{\Bigl(1-\frac{ \exp\left[-m_{1}r\right]}{\left(3+2\omega\right)}
 - \frac{\epsilon\tau_{3} r^3}{12GM_s}\Bigr)} {\Bigl(1+\frac{\exp\left[-m_{1} r\right]}{\left(3+2\omega\right)} 
 +\frac{\epsilon\tau_{3} r^3}{12GM_s}\Bigr)}
 \label{gammaPhi}
\eeq

We analyze the constraints on the parameters appearing in the above expression in section III. At present we consider a second case of the non-minimal coupling function. But before that it is worth mentioning
one interesting situation  when the potential $V(\Phi)$ in the action  (\ref{action}) is a constant. This  case is interesting because it gives rise to the Chaplygin Gas equation of state $p = -\frac{A}{\rho}$  where $p$ is the pressure and $\rho$ is the energy density and $A$ being an arbitrary constant. This has been studied extensively in cosmology \cite{chaplygin}. In this case the  $ h_{00}$ and $h_{ij}$ are of the form:
\begin{align}
h_{00}=& \frac{2GM_s}{\Phi_0 r} \Bigl(1+\frac{ 1}{\left(3+\tau_3 \eta^2 V_c \Phi_0\right)}\Bigr)\nonumber\\
& +\Bigl[\frac{\tau_3 V_c}{6 \Phi_0}-\frac{\tau_3 V_c}{3 \Phi_0 \left(3+\tau_3 \eta^2 V_c \Phi_0\right)\Phi_0}\Bigr]r^2
\end{align}and
\begin{align}
h_{ij}=& \frac{2GM_s \delta_{ij}}{\Phi_0 r} \Bigl(1-\frac{ 1}{\left(3+\tau_3 \eta^2 V_c \Phi_0\right)}\Bigr)\nonumber\\
& -\Bigl[\frac{\tau_3 V_c}{6 \Phi_0}-\frac{\tau_3 V_c}{3 \Phi_0 \left(3+\tau_3 \eta^2 V_c \Phi_0\right)\Phi_0}\Bigr]r^2\delta_{ij}
\end{align}

\noindent
where $V_{c} = V(\Phi) = constant$. In the case where, the terms inside the square bracket cancel each other,  the solutions are exactly identical to the massless Brans-Dicke case with the identification $\tau_{3} \eta^2 V_{c} \Phi_{0} = 2\omega$ where $\omega$ is the Brans-Dicke parameter. The parameter $\gamma$ in this case is given by 
\beq
\gamma = \frac{\tau_{3} \eta^{2} V_{c} \Phi_{0}+2}{\tau_{3}\eta^2 V_{c} \Phi_{0}+4}
\eeq

 \begin{figure*}[t]
\centering
\begin{center}
$\begin{array}{@{\hspace{-0.10in}}c@{\hspace{0.0in}}c}
\multicolumn{1}{l}{\mbox{}} & \multicolumn{1}{l}{\mbox{}} \\
[-0.2in] \epsfxsize=3.3in \epsffile{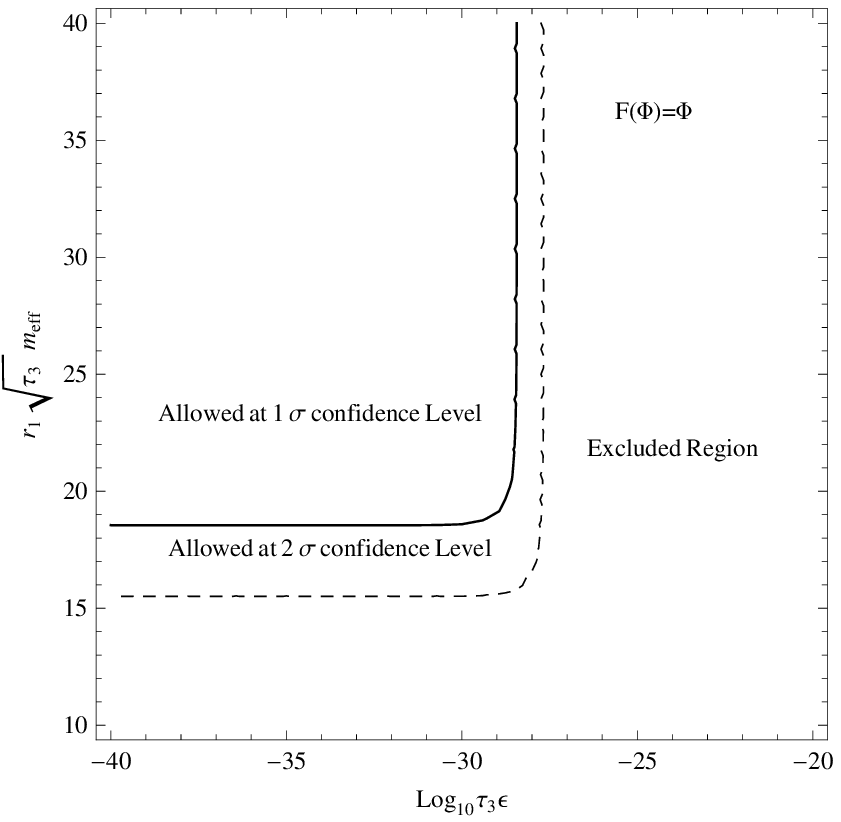} & \epsfxsize=3.3in
\epsffile{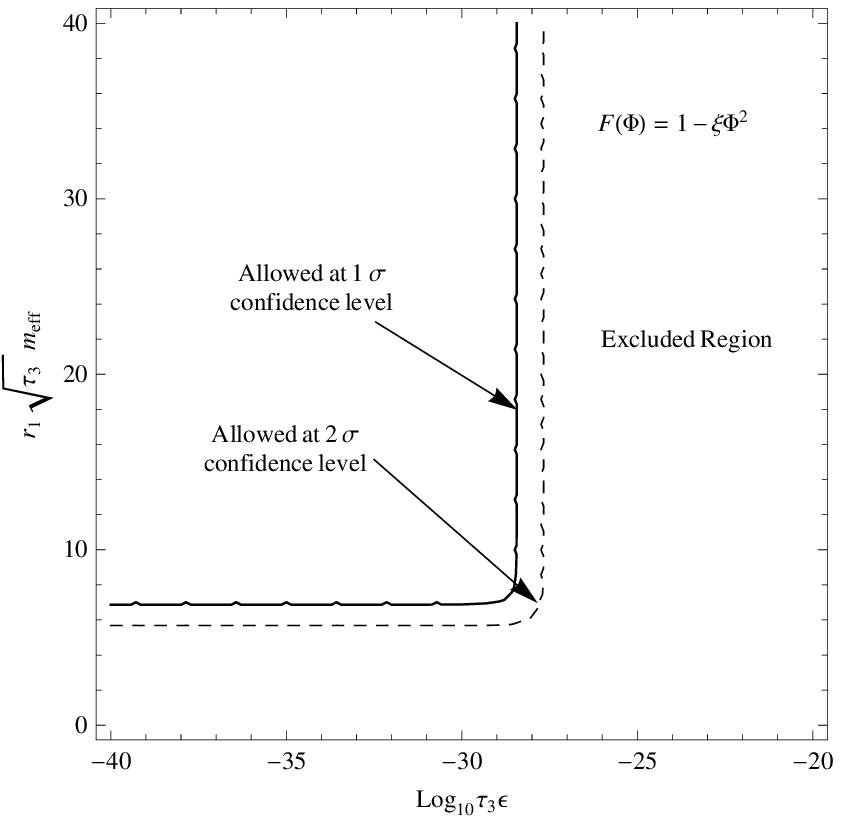} \\
\end{array}$
\end{center}
\vspace{0.0cm} \caption{\small ({\it left}). The observationally allowed regions for the parameters $\tau_{3}\epsilon$  and $\tau_{3} m_{eff}^2 r_1^2$ at $1\sigma$ 68\% confidence level and $2\sigma$ 95\% confidence level for the coupling factor $F(\Phi)=\Phi$ taking $\Phi_{0}=1$ . ({\it right}). Same as ({\it left}) but for the coupling $F(\Phi)=(1-\xi \Phi^2)$  with $\Phi_{0}=1$and $\xi=1/6$.  } 
 
\end{figure*}
 \section{Case(ii)  $\!\!  F(\Phi) = 1- \xi \Phi^2$}
  We consider the  case of conformal coupling given by
\beq
 F(\Phi) = 1- \xi \Phi^2
\eeq 

\noindent
with $\xi = \frac{1}{6}$. 

\noindent
We proceed in a similar way as in case(I), but with a different gauge choice for simplifying the analysis :
\beq
 h^{\mu}_{\nu},_{\mu} -\frac{1}{2} h^{\mu}_{\mu},_{\nu} = \frac{F'(\Phi_0)}{F(\Phi_0)}\psi,_{\nu}.
\eeq

\noindent
 As in the previous case, we define  the variable  $\chi_{\mu\nu}=h_{\mu\nu}-\eta_{\mu\nu}\frac{h}{2}-\eta_{\mu\nu} \frac{F'(\Phi_0)}{F(\Phi_0)}\psi$ in order to write the equations in a simplified form. With this equations (\ref{Gmu}) and (\ref{field}) become

\beq
-\frac{F(\Phi_0)}{2}\nabla^2 \chi_{\mu\nu} = 8\pi GT_{\mu\nu}-\frac{\tau_3}{2} \epsilon\eta_{\mu\nu}
\eeq

\beq
\nabla^2\psi-\frac{\alpha^2}{\beta^2}\psi- 
\frac{4\tau_3\epsilon \xi \Phi_0}{\beta^2}
= -\frac{16\pi G\rho \xi \Phi_0}{\beta^2},
\eeq 

\noindent 
where $\alpha^2=\tau_3 F(\Phi_0)m_{eff}^2$ and $\beta^2=\left[ \tau_3 \eta^2 \epsilon F(\Phi_0)+12\xi^2 \Phi_0^2\right]$. For the energy density, $\rho = M_s\delta(r)$, the solution of the above equations are
\beq
\psi = -\frac{4GM_s \xi \Phi_0}{\beta^2 r}\exp\left[-\sqrt{\frac{\alpha^2}{\beta^2}}r\right] -\frac{ 4\epsilon \xi \Phi_0}{F(\Phi_0)m_{eff}^2}
\eeq
\beq
\chi_{00} = \frac{4 G M_s}{F(\Phi_0)r} - \frac{\tau_3 \epsilon}{6 F(\Phi_0)} r^2
\eeq 
\beq
\chi_{ij} = \frac{\tau_3 \epsilon}{6 F(\Phi_0)} r^2 \delta_{ij}.
\eeq
In term of $h_{00}$ and $h_{ij}$, the solutions become

\ber
h_{00} =   \frac{2 G M_s}{F(\Phi_0)r} \Bigg(1+ \frac{4\xi^2 \Phi_0^2 \exp\left[-\sqrt{\frac{\alpha^2}{\beta^2}} r\right]}{\beta^2}\Bigg) \nonumber\\
+ \frac{8\xi^2 \Phi_0^2 \epsilon}{F(\Phi_0)^2 m_{eff}^2} + \frac{\tau_3 \epsilon}{6 F(\Phi_0)} r^2
\eer

\ber
h_{ij} =  \frac{2 G M_s\delta_{ij}}{F(\Phi_0)r}\Bigg(1- \frac{ 4\xi^2 \Phi_0^2}{\beta^2}
 \exp \left[-\sqrt{\frac{\alpha^2}{\beta^2}} r\right] \Bigg)\nonumber\\
-\frac{8\xi^2 \Phi_0^2 \epsilon}{F(\Phi_0)^2 m_{eff}^2}\delta_{ij} - \frac{\tau_3 \epsilon}{6 F(\Phi_0)}\delta_{ij} r^2
\eer

\noindent
Likewise the linear coupling, here also we get both the constant term and the term proportional to $r^2$ together with the usual exponential correction. The constant term can again be removed with a suitable coordinate transformation $x^{\mu} \rightarrow x^{\mu} + \zeta^{\mu}$ where $\zeta^{\mu} = 4\frac{\xi^2\Phi_{0}^2\epsilon}{F(\Phi_{0})^2 m_{eff}^2}$ which keeps the gauge condition (27) unchanged. 

\noindent
The effective gravitational constant $G_{eff}$ in this case, can be written as 
\ber
G_{eff}= \frac{G}{F(\Phi_0)}\Bigg(1+ \frac{4\xi^2 \Phi_0^2}{\beta^2} 
\exp{\left[-\sqrt{\frac{\alpha^2} 
{\beta^2}}r\right]}\nonumber\\
+\frac{\epsilon\tau_{3} r^3}{12GM_s}\Bigg)
\eer
The Post-Newtonian Parameter $\gamma$ for this case is found to be
\beq
\gamma=  
\frac{1- \frac{4\xi^2 \Phi_0^2}{\beta^2}\exp\left[-\sqrt{\frac{\alpha^2}{\beta^2}} r\right]
-\frac{\epsilon\tau_{3} r^3}{12GM_s}}{1+ \frac{4 \xi^2 \Phi_0^2 }{\beta^2} \exp\left[-\sqrt{\frac{\alpha^2}{\beta^2}}r\right]
 +\frac{\epsilon\tau_{3} r^3}{12GM_s}}
\label{gammaPhiPhi} 
\eeq

\section{Observational Constraints}

In order to constrain the model parameters, we use the recent measurement of  time delay $\Delta \tau$  of the radio waves transmission near the solar conjunction by the Cassini spacecraft \cite{Bertotti(2003)}.

The delay in time taken for the signal to travel the round trip distance between a ground based antenna and a reflector placed in a spacecraft is produced due the gravitational field of the Sun. For the standard Einstein's gravity, it has the form of 
\beq
\Delta \tau =2(1+\gamma) GM_{s}ln
\left(\frac{4r_{1}r_{2}}{b^2}\right)
\eeq
where G is the gravitational constant, b is the impact
parameter and $r_{1}$ and $r_{2}$ are distances of the ground base antenna and the spacecraft respectively from the Sun. Here, $\gamma$ is a constant and its measured value by Cassini mission is $\gamma_{obs} =1 + (2.1 \pm 2.3)\times 10^{-5}$ \cite{Bertotti(2003)}. But the expression (37)  is only valid for constant $\gamma$. If $\gamma$ is not a constant, as in our case ( in fact $\gamma$ is not a constant for most of the modified gravity models), it is wrong to use the expression for $\gamma$ [as in equation (22) and (36)] together with the above mentioned constraint from Cassini mission to put bound on the model parameters.   In that case, one has to calculate the actual time delay $\Delta\tau$ for the model.  Doing this in our model, we get $\Delta\tau$ for $F(\Phi)=\Phi$ case as 
\ber
\Delta \tau = \frac{4GM_s}{\Phi_0}ln\left[\frac{(a_T+r_{1})(a_R+r_{2})}{b^2}\right]
-\frac{2GM_s}{\Phi_0}\frac{(a_R+a_T)}{r_{1}}\nonumber\\
\left(\frac{1}{3+2w}\exp[-\sqrt{\frac{\tau_3 \Phi_0}{3+2w}}m_{eff}r_{1}]+\frac{\epsilon\tau_{3} r^3}{12GM_s}\right)\nonumber\\
\eer
Here, $a_{T}=\sqrt{r_{1}^2-b^2}$ and $a_{R}=\sqrt{r_{2}^2-b^2}$.
For the conformal coupling case $F(\Phi)=(1-\xi \Phi^2)$, we get similarly as
\begin{align}
& \Delta \tau = \frac{4GM_s}{(1-\xi\Phi_0^2)}ln\left[\frac{(a_T+r_1)(a_R+r_{2})}{b^2}\right] \nonumber\\
&-\frac{2GM_s}{(1-\xi\Phi_0^2)}\frac{(a_R+a_T)}{r_1}\nonumber\\
&\left(\frac{4\xi^2\Phi_{0}^2}{\beta^2}\exp[-\sqrt{\frac{\alpha^2}{\beta^2}}r_1] +\frac{\epsilon\tau_{3} r^3}{12GM_s}\right)\nonumber\\
\end{align}

\noindent

The observation bound on $\Delta \tau_{obs} $ is \cite{Bertotti(2003)}
\beq 
\Delta \tau_{obs}> 2.6042\times 10^{-4}sec
\label{delay1}
\eeq
\beq
\Delta \tau_{obs} >2.60417\times 10^{-4}sec 
\label{delay2}
\eeq
at the $1\sigma$ and $2\sigma$ levels respectively.   

\noindent

Using equations (38) and (39), we now put constraints on the model parameters using the bound on $\Delta\tau$ mentioned above.

In figure 1 ({\it left}), we show the constraint on the parameter $\tau_{3}\epsilon$ and $m_{eff}r_{1}\sqrt{\tau_{3}}$ for the case of linear coupling. For this we assume a particular value for $\Phi_{0} = 1$. One should keep in mind that these two parameters are related to the energy scale and the mass scale of the potential. So the allowed region actually constraints the shape of the potential around its extremum. 

Similarly in Figure 1 ({\it right}), we show the same constraint  for the conformal coupling case with the same value of $\Phi_{0}$. Comparing these two figures, one can say that in the conformal coupling case, smaller mass for  the scalar field  is allowed for the same range of scalar field energy density. 

\section{Conclusion}

In this paper, we analyze the Solar system constrain on the scalar tensor theories having a nonstandard action for the scalar field. We consider the linear (same as BD case) as well as the conformal couplings case. In both the cases, the Newtonian potential is corrected by a term which goes as square of the distance, typical in presence of a cosmological constant.  In our case, this term is due to nonzero scalar field energy density at the extremum. This is in addition to the well known exponentially suppressed correction which depends on the effective mass of the scalar field.

Next we calculate the PPN parameter $\gamma$ in our model and have shown that together with the usual exponential correction term, there is another term which has $r^2$ dependence. The exponential correction term comes together with a factor which is similar to $(2\omega+3)$ as in BD case. But unlike the BD case where $\omega$ is an independent parameter, here it depends upon the potential parameters.

We also calculate the actual time delay for the signal to  travel the round trip distance between a ground based antenna and a reflector placed in a spacecraft due to gravitational field of the Sun for our model. Using the measurement of this time delay by the Cassini mission, we find the allowed region for our parameters $\tau_{3} \epsilon$  and $\tau_{3}m_{eff}^2 r_1^2$ which essentially control the shape of the potential around its extrema.  Hence local gravity tests like time delay meaurements can be useful in constraining the shape of the potentials for scalar fields having nonstandard actions and nonminimally coupled to the gravity sector.

\section{Acknowledgment}
A.A.S and N.C. Devi acknowledge the  financial support provided by the
University Grants Commission, Govt. Of India, through major
research project grant (Grant No:33-28/2007(SR)). N.C. Devi acknowledges the hospitality provided by the Harish-Chandra Research Institute, Allahabad, India where part of the work has been done. The authors are grateful to Ashoke Sen, Debashis Ghoshal and M. Sami for the comments and suggestions. The author are also thankful to the anonymous referee for his valuable comments and suggestions which improves the clarity of the paper.

\end{document}